\documentclass[fleqn,twoside]{article}
\usepackage{espcrc2}
\def\photonatomrightt{\begin{picture}(3,1.5)(0,0)
                               \put(0,-0.75){\tencirc \symbol{2}}
                               \put(1.5,-0.75){\tencirc \symbol{1}}
                               \put(1.5,0.75){\tencirc \symbol{3}}
                               \put(3,0.75){\tencirc \symbol{0}}
                     \end{picture}
                    }

\def\photonrightthalf{\begin{picture}(15,1.5)(0,0)
                    \multiput(0,0)(3,0){5}{\photonatomrightt}
                 \end{picture}
                }

\def\fermionrighthalf{\begin{picture}(15,1)(0,0)  
                            \put(0,0){\line(1,0){7.5}}
                          \put(7.5,0){\line(1,0){7.5}}
                      \end{picture}
                     }
\def\fermionul{\begin{picture}(15,15)(0,0)
                        \put(0,0){\vector(-1,1){7.5}}
                        \put(-7.5,7.5){\line(-1,1){7.5}}
                  \end{picture}
                 }

\def\fermionur{\begin{picture}(15,15)(0,0)
                        \put(-15,-15){\vector(1,1){7.5}}
                        \put(-7.5,-7.5){\line(1,1){7.5}}
                  \end{picture}
                 }

\def\gaugebosonright{\begin{picture}(30,1)(0,0)
                            \put(0,0){\line(1,0){0.75}}
                            \multiput(2.25,0)(3,0){9}{\line(1,0){1.5}}
                            \put(29.25,0){\line(1,0){0.75}}
                     \end{picture}
                    }

\def\gaugebosonurhalf{\begin{picture}(15,15)(0,0)
                            \put(0,0){\line(1,1){15.0}}
                  \end{picture}
                 }

\def\gaugebosondrhalf{\begin{picture}(15,15)(0,0)
                            \put(0,0){\line(1,-1){15}}
                  \end{picture}
                 }
\def\gaugebosondrhalff{\begin{picture}(7.5,7.5)(0,0)
                            \put(0,0){\line(1,-1){7.5}}
                  \end{picture}
                 }

\newenvironment{Feynman}[3]{\begin{center}
                            \setlength{\unitlength}{#3 mm}
                            \begin{picture}(#1)(#2)
                            \thicklines
                           }{\end{picture} \end{center}}

\usepackage{graphicx}

\newcommand{\AmS}{{\protect\the\textfont2
  A\kern-.1667em\lower.5ex\hbox{M}\kern-.125emS}}

\title{Higgs physics at a high luminosity 
$\rm e^+e^-$  linear collider}

\author{Andr\'e Sopczak\thanks{On behalf of the ECFA/DESY Higgs boson study group
        with contributions from US and Asian Higgs study groups.}, 
       {Lancaster University, UK}
}

\begin{document}
\begin{titlepage}
\def\thefootnote{\fnsymbol{footnote}}       

\begin{center}
\mbox{ } 

\end{center}
\vskip -1.0cm
\begin{flushright}
\Large
\vspace*{-3cm}
\mbox{\hspace{10.2cm} hep-ph/0209372} \\
\mbox{\hspace{12.0cm} September 2002}
\end{flushright}
\begin{center}
\vskip 2.0cm
{\Huge\bf
Higgs physics 
\smallskip
at a high luminosity 
\boldmath$\rm e^+e^-$\unboldmath\ linear collider}
\vskip 1cm
{\LARGE\bf Andr\'e Sopczak}\\
\smallskip
\Large Lancaster University, UK

\vskip 2.cm
\centerline{\Large \bf Abstract}
\end{center}

\vskip 3.5cm
\hspace*{-1cm}
\begin{picture}(0.001,0.001)(0,0)
\put(,0){
\begin{minipage}{17.5cm}
\Large
\renewcommand{\baselinestretch} {1.2}
Since the TESLA Technical Design Report (TDR) was published in 2001, the
physics programme of an electron-positron linear collider (LC)
has been further developed and a wide consensus has been reached on the
physics case and the need for a high luminosity LC with center-of-mass
energy up to about 1 TeV as the next worldwide high-energy physics project.
The study of the Higgs boson properties represents a significant part of
this physics programme. New studies demonstrate that a LC providing 
1000~fb$^{-1}$ of data at center-of-mass  energies of at least 500 GeV, 
is an excellent
Higgs boson analyzer for a wide range of masses. 
A summary of preliminary
results of these studies and their implications for identifying the
nature of the Higgs sector and for constraining the parameter space
of extended models is given. The focus is on recent developments and
the relation to the LHC is addressed.
\renewcommand{\baselinestretch} {1.}

\normalsize
\vspace{2cm}
\begin{center}
{\sl \large
Presented at the 
International Conference on High Energy Physics (ICHEP02), \\
Amsterdam, July 2002
\vspace{-3cm}
}
\end{center}
\end{minipage}
}
\end{picture}
\vfill

\end{titlepage}


\newpage
\thispagestyle{empty}
\mbox{ }
\newpage
\setcounter{page}{1}

\begin{abstract}
\vspace*{-0.2cm}
Since the TESLA Technical Design Report (TDR) was published in 2001, the
physics programme of an electron-positron linear collider (LC)
has been further developed and a wide consensus has been reached on the
physics case and the need for a high luminosity LC with center-of-mass
energy up to about 1 TeV as the next worldwide high-energy physics project.
The study of the Higgs boson properties represents a significant part of
this physics programme. New studies demonstrate that a LC providing 
1000~fb$^{-1}$ of data at center-of-mass  energies of at least 500 GeV, 
is an excellent
Higgs boson analyzer for a wide range of masses. 
A summary of preliminary
results of these studies and their implications for identifying the
nature of the Higgs sector and for constraining the parameter space
of extended models is given. The focus is on recent developments and
the relation to the LHC is addressed.
\vspace{1pc}
\vspace*{-0.8cm}
\end{abstract}

\maketitle

\section{INTRODUCTION}
\vspace*{-0.1cm}
A linear collider of at least 500~GeV and a total luminosity of at least
1000~fb$^{-1}$ has much potential for studying Higgs bosons and 
understanding 
the electroweak symmetry breaking and mass generation. 
Already at the CERN $\rm e^+e^-$ collider, LEP, enormous progress 
for Higgs boson searches has been made.
At \mbox{LEP-1~\cite{lep1}} many search channels were almost background 
free and also at \mbox{LEP-2}~\cite{lepwg} the sensitivity exceeded 
expectations,
leading to a Standard Model (SM) Higgs boson mass limit of 114~GeV at 95\% CL.
A small indication at LEP of a 115~GeV SM Higgs boson could also be 
interpreted in extended models~\cite{as}.

During the last 10 years, LC studies have evolved from discovery studies
to studies of precision measurements. Recent milestones were set with the
TESLA TDR~\cite{tdr}, 
the Snowmass Study~\cite{snowmass} and 
the international workshop on LC in Korea~\cite{korea}.

This review will focus on new results and developments.
First, I will address the SM physics and
discuss how precisely a future LC can determine 
the Higgs boson production mechanism. 
Indirect and direct branching ratio measurements are reviewed. 
Then the characterization of the Higgs boson potential, which 
will contribute to establishing the underlying mechanism of mass generation,
is addressed.
Higgs bosons could also be produced via Higgs-strahlung off
top quarks. 
In the general two Higgs doublet model (2HDM) charged Higgs bosons are
prominent. 
Various methods to determine the ratio of the vacuum expectation values
of the two doublets, $\tan\beta$, are discussed.
In the framework of the Minimal Supersymmetric extension of the SM (MSSM) 
or beyond, invisibly decaying Higgs bosons could be produced and their 
properties measured. 
Furthermore, the measurement of the Higgs boson parity is discussed as
well as important possibilities to distinguish Higgs boson models.
For several LC studies the relation to the LHC potential is addressed.
The importance of beam polarization and the option of a $\gamma\gamma$
collider are reviewed elsewhere~\cite{gudi,stefan}.
Future LC Higgs studies will concentrate on
more detailed detector simulations reflecting the progress in detector 
technologies,
the second phase of a LC with center-of-mass energies up to 1~TeV for 
TESLA,
and new theoretical developments, such as extra dimensions.

\section{STANDARD MODEL PHYSICS}

\subsection{Higgs boson production mechanism}
\begin{figure}[htb]
\includegraphics[width=\columnwidth]{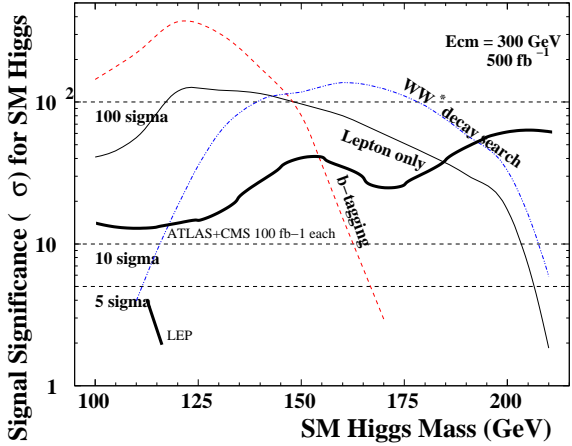}
\vspace*{-1.4cm}
\caption{SM Higgs boson signal significance.}
\vspace*{-0.8cm}
\label{fig:smmass}
\end{figure}

The expected SM Higgs production rate
has a very large significance over the background
($\sigma\equiv N_{\rm sig}/\sqrt{N_{\rm bg}}$)~\cite{yamashita}.
More than $100\sigma$ is obtainable in the $\rm H\rightarrow b\bar b$ 
decay channel.
For heavier Higgs bosons, the WW decay mode takes over.
In relation to the LHC~\cite{gianotti}, a LC has a much larger
sensitivity in the lower mass range, 
while the LHC can probe heavier Higgs bosons.
At about 115~GeV mass, the LEP sensitivity reduces from 
about $4\sigma$ to $2\sigma$~\cite{lepwg}.
Figure~\ref{fig:smmass} (from~\cite{yamashita}) shows the SM Higgs boson 
detection significance.

\begin{figure}[htb]
\vspace*{-1.7cm}
\includegraphics[width=1.1\columnwidth]{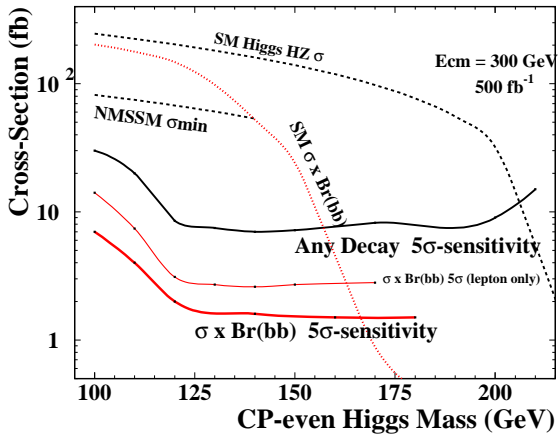}
\vspace*{-1.9cm}
\caption{Expected production cross sections (dashed lines)
         and simulated experimental sensitivities (solid lines).}
\label{fig:xsec}
\end{figure}

The sensitivity to the general production cross section has been
studied~\cite{yamashita}.
The expected SM production cross section is much larger than the 
$5\sigma$ sensitivity to the cross section over a wide mass range. 
For lower masses the Higgs boson
decay mode into a pair of b-quarks gives the largest sensitivity.
Figure~\ref{fig:xsec} (from~\cite{yamashita}) 
shows also the cross section sensitivity for
any Higgs boson decay mode, 
where the Higgs boson mass 
is reconstructed as the recoiling mass to the Z decay products. 
Even for models beyond the SM, the sensitivity to the
cross section is much better than the expected minimal 
production cross section.

\begin{figure}[htb]
\vspace*{-0.5cm}
\includegraphics[width=\columnwidth, bbllx=284pt,bblly=479pt,bburx=515pt,bbury=631pt,clip=]{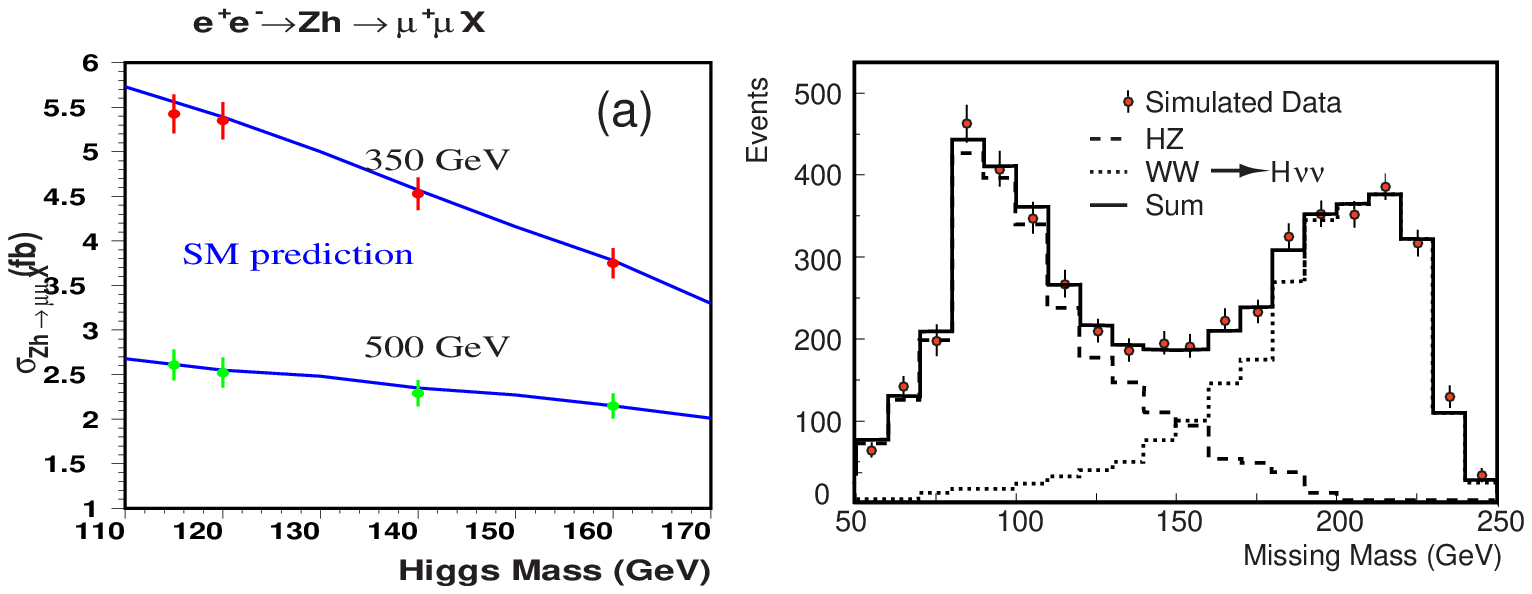}\\
\vspace*{-1.4cm}
\caption{Expected missing mass distributions for Higgs-strahlung 
         and WW fusion. Their sum is shown in comparison with 
         simulated data.}
\label{fig:production}
\end{figure}

The Higgs boson will be produced via Higgs-strahlung and 
WW fusion by the two processes 
$$\rm e^+e^-\rightarrow HZ \rightarrow H\nu\bar\nu \rightarrow b\bar b\nu\bar\nu~~~and
$$
$$\rm e^+e^-\rightarrow WW\nu\bar\nu \rightarrow \nu\bar\nu H\rightarrow \nu\bar\nu b\bar b.
$$
These production mechanisms can be distinguished by fitting the missing
mass distribution~\cite{meyerdesch} as shown in Fig.~\ref{fig:production}
(from~\cite{rick}).
Higgs-strahlung gives a shape like the dashed line, while Higgs boson
fusion is represented by the dotted line.
Their sum is given by the solid line and the simulated data is shown with
error bars.

\subsection{Indirect and direct branching ratio measurements}
A LC will perform very high precision measurements of the Higgs boson
branching fractions. The underlying production process is the 
Higgs-strahlung, 
$$\rm e^+e^-\rightarrow HZ\rightarrow H\ell^+\ell^-,$$ 
where the associated Z boson decays into a lepton pair.
Two methods are discussed to determine the Higgs boson branching ratio. 

In the indirect method, the inclusive production cross section as the 
product of the Higgs boson production cross section 
and the branching fraction of the
Z into leptons is determined: 
$$\sigma_{\rm inc} = \sigma_{\rm HZ}
            BR(\rm Z\rightarrow \ell^+\ell^-).$$
This measurement is independent of the
Higgs boson decay mode. The mass recoiling to the lepton pair corresponds 
to the Higgs boson mass. An individual Higgs boson decay cross section
is measured, which is the product of the Higgs boson production cross
section and the Higgs and Z boson decay branching fractions:
\vspace*{-1mm}
$$\sigma({\rm X}) = \sigma_{\rm HZ}
            BR({\rm Z\rightarrow Y}) 
            BR({\rm H\rightarrow X}).$$
By taking the ratio
of both cross sections the Higgs branching ratio can be determined, since
the LEP experiments measured the Z decay branching fractions with
high precision.

\begin{figure}[htb]
\vspace*{-4.0cm}
\includegraphics[width=1.0\columnwidth]{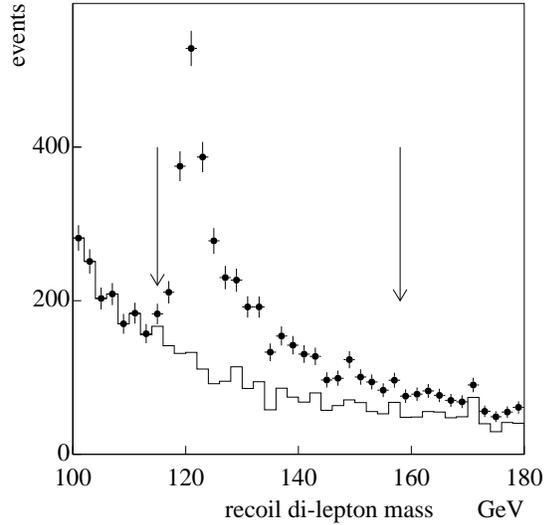}
\vspace*{2.1cm}
\caption{Simulated events for a direct determination of the 
         Higgs boson decay branching ratios.}
\label{fig:br}
\vspace*{-0.7cm}
\end{figure}

The Higgs boson branching ratios can also be determined directly 
from the HZ($\rm Z\rightarrow \ell^+\ell^-$) 
event sample~\cite{brient}.\,The Higgs boson\,mass is\,recon\-structed 
as the recoiling mass of the lepton pair: 
$$m_{\rm H}=m_{\ell^+\ell^-}^{\rm recoil}.$$
\vspace*{-1mm}
The simulation was performed for $\sqrt s=360$~GeV and 
${\cal L}=500$~fb$^{-1}$~\cite{brient}.
Figure~\ref{fig:br} shows this event sample which is enriched with Higgs
bosons by the indicated cuts.
The expected signal distribution 
is given with
error bars and the expected background as a histogram.
In this event sample individual Higgs boson decay modes are selected. 
The resulting
precision on the Higgs boson decay branching ratios 
as well as a preliminary combination of both methods~\cite{brient} 
are listed in Table~\ref{table:br}.

\begin{table}
\caption{Expected\,precision\,on\,branching\,ratios\,(in\,\%)\,for\,a 
         120\,GeV\,Higgs\,boson\,from\,the\,direct\,method\,and\,a 
         preliminary\,combination\,with\,the\,indirect\,method.}
\label{table:br}
\renewcommand{\arraystretch}{1.2} 
\begin{center}
\begin{tabular}{cccc}
Decay    &SM $BR$  &$\Delta BR_{\rm d} /BR_{\rm d}$&$\Delta BR_{\rm c} /BR_{\rm c}$ \\\hline
bb       & 68       &1.9 & 1.5    \\
$\tau\tau$ & 6.9    &7.1 & 4.1  \\
cc       & 3.1      &8.1 & 5.8 \\
gluons   & 7.0      &4.8 & 3.6 \\
$\gamma\gamma$&0.22 &35  & 21 \\ 
WW$^\star$ & 13     &3.6 & 2.7     
\vspace*{-0.8cm}
\end{tabular}
\end{center}
\end{table}

In relation to the LHC, a LC will achieve a much higher precision
and will cover all decay modes. 
This would also allow precision testing of the fundamental
relation between the Yukawa coupling and the Higgs boson mass:
\vspace*{-1mm}
$$g_{\rm Hff}\propto m_{\rm f}.
\vspace*{-3mm}
$$

\subsection{Mass and width determination}
\begin{figure}[htb]
\includegraphics[width=\columnwidth]{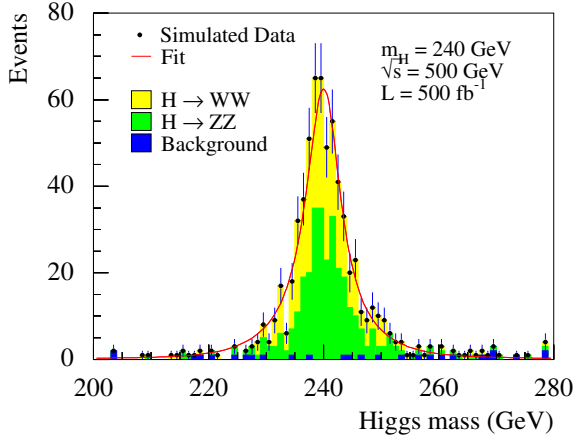}
\vspace*{-1cm}
\caption{Mass and width determination.}
\label{fig:mass}
\vspace*{0.2cm}
\end{figure}

Further properties of the Higgs boson, like its mass and decay 
width, can be determined with high precision.
A study of a 240~GeV SM Higgs boson in the reactions
$$\rm e^+e^-\rightarrow HZ\rightarrow WWZ~~~and$$
$$\rm e^+e^-\rightarrow HZ\rightarrow ZZZ$$
has been performed~\cite{meyer}. A very clear signal with only
a few background events is expected. 
Figure~\ref{fig:mass} shows the expected signal and background mass
distribution.
The fit of the reconstructed
mass peak leads to a mass resolution of 0.08\% and an 11\% error
on the total decay width.

\begin{figure}[htb]
\vspace*{-0.3cm}
\includegraphics[width=\columnwidth]{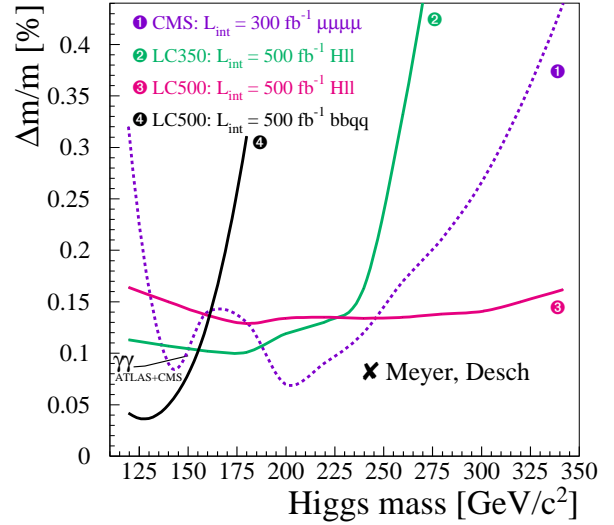}
\vspace*{-1.5cm}
\caption{Comparison of different methods to determine the Higgs boson
         mass.}
\label{fig:masscombi}
\vspace*{-0.1cm}
\end{figure}

This result is shown in comparison with results from other reactions
on the determination of the Higgs boson mass~\cite{drollinger}.
Figure~\ref{fig:masscombi} shows results from a CMS study (curve 1)
where the Higgs boson decays into a pair of Z bosons, which subsequently
decay into muon pairs. The sensitivity reduction near 160 GeV is due
to dominant decays into W pairs.
The curves 2 and 3 show extrapolated results according to the cross 
section and branching ratio expectations for a LC operation at 350 and 
500~GeV.
In the low mass region indirect methods can be applied (curve 4)
and for very low masses, the LHC has high sensitivity in the photon decay 
mode.

\subsection{Characterization of the Higgs boson potential}
\begin{figure}[htb]
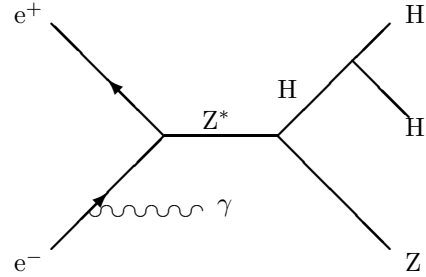

\vspace*{-0.1cm}
\begin{center}
\begin{Feynman}{60,33}{1,27}{1.0}
\put(25,40){\fermionul}        \put(5,22){${\rm e^-}$}
\put(25,40){\fermionur}        \put(5,55){${\rm e^+}$}
\put(15,30){\photonrightthalf} \put(32,30){$\gamma$}
\put(25,40){\fermionrighthalf}  \put(30,41){${\rm Z^*}$}
\put(40,40){\gaugebosonurhalf} \put(57,55){${\rm H}$}
\put(40,40){\gaugebosondrhalf} \put(57,22){${\rm Z}$}
\put(50,50){\gaugebosondrhalff} \put(57,40){${\rm H} $}
                                \put(40,45){${\rm H} $}
\end{Feynman}
\end{center}
\vspace*{-0.7cm}
\caption{Higgs boson self-coupling reaction.}
\label{fig:hhh}
\end{figure}

A LC will be able to measure fundamental properties of the Higgs boson
potential. The Higgs boson can decay into a pair of Higgs bosons:
$$\rm e^+e^-\rightarrow HZ\rightarrow HHZ.$$
Figure~\ref{fig:hhh} illustrates this Higgs boson production reaction.

\begin{figure}[htb]
\includegraphics[width=\columnwidth]{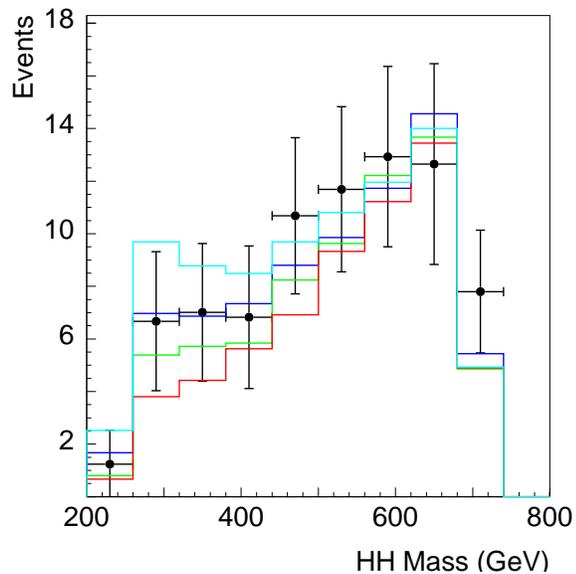}
\vspace*{-1.0cm}
\caption{Reconstructed invariant mass of the HH pairs.}
\label{fig:potential800}
\end{figure}

\begin{figure}[htb]
\includegraphics[width=\columnwidth]{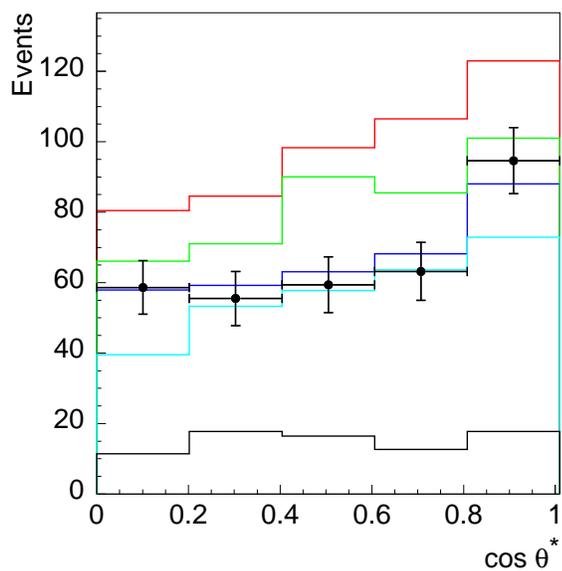}
\vspace*{-1.0cm}
\caption{Reconstructed angle between the H and HH directions.}
\label{fig:potential3000}
\end{figure}

The self-coupling interaction can probe the shape of the Higgs boson
potential 
through the relation
$$g_{\rm HHH} = 3m_{\rm H}^2/2v,$$
where $v=246$~GeV.
A sensitivity of $\Delta g/g=29$\% was obtained for
$m_{\rm H} =120$~GeV, $\sqrt s=800$~GeV and 
${\cal L}=1000$~fb$^{-1}$~\cite{p_gay,potential}.
Figure~\ref{fig:potential800} shows the precision on 
the Higgs boson self-coupling where the lines indicate 
$g_{\rm HHH}/g_{\rm HHH}^{\rm SM} =1.25,~1.00,~0.75,~0.50$.
Higher sensitivity of $\Delta g/g=7$\% could be reached for 
a LC with $\sqrt s=3$~TeV and ${\cal L}=5000$~fb$^{-1}$ as 
shown in Fig.~\ref{fig:potential3000}.

\subsection{Higgs-strahlung from top quarks}
\begin{figure}[htb]
\vspace*{-1cm}
\begin{center}
\includegraphics[width=0.7\columnwidth]{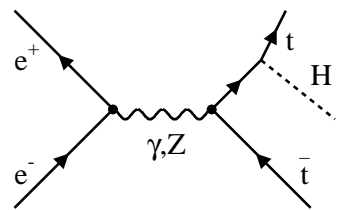}
\end{center}
\vspace*{-1cm}
\caption{$\rm t\bar tH$ production process.}
\label{fig:tth}
\end{figure}

\begin{figure}[htb]
\includegraphics[width=\columnwidth]{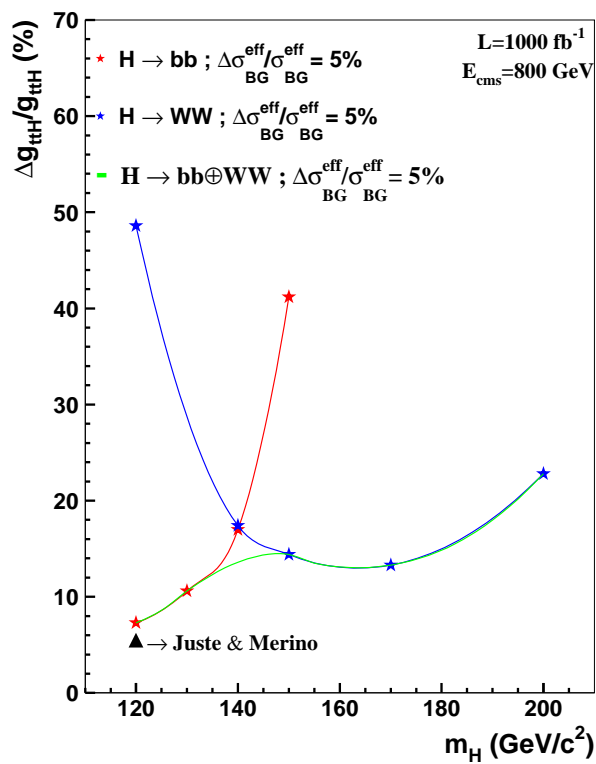}
\vspace*{-1cm}
\caption{Expected precision on the Yukawa coupling $g_{\rm ttH}$.}
\label{fig:yukawa}
\end{figure}

Owing to the strong coupling of Higgs bosons to top
quarks, the radiation of Higgs boson off top quarks is
a possible production mechanism (Fig.~\ref{fig:tth}). 
The decay modes 
involving a pair of b quarks and W bosons 
were studied in the reactions~\cite{gay}
$$\rm e^+e^-\rightarrow t\bar tH\rightarrow t\bar t b\bar b~~~and$$
$$\rm e^+e^-\rightarrow t\bar tH\rightarrow t\bar t WW.$$
A challenge is the precision determination of the background
to a level of 5\% uncertainty, leading to
$$\Delta g_{\rm ttH}/g_{\rm ttH}=7.5\%~{\rm to}~20\%.$$
Figure~\ref{fig:yukawa} shows the resulting precision for the $\rm b\bar b$
and WW decay modes, as well as their statistical combination
as a function of the SM Higgs boson mass.
In addition, a previous study at 120~GeV with slightly higher sensitivity 
is indicated~\cite{aurelio}.

\clearpage
\section{BEYOND THE STANDARD MODEL}

\subsection{Charged Higgs bosons}
\begin{figure}[htb]
\includegraphics[width=\columnwidth]{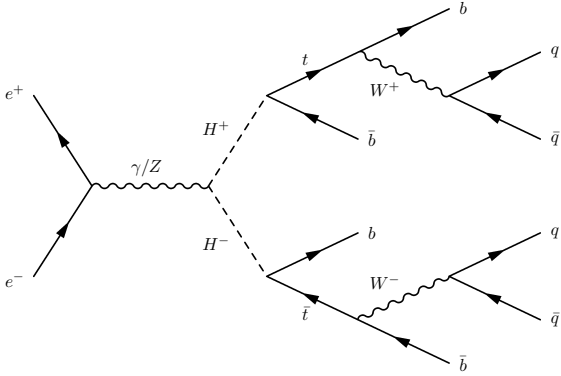}
\vspace*{-1cm}
\caption{Charged Higgs boson production and decay.}
\label{fig:hphm}
\end{figure}

\begin{figure}[htb]
\includegraphics[width=\columnwidth]{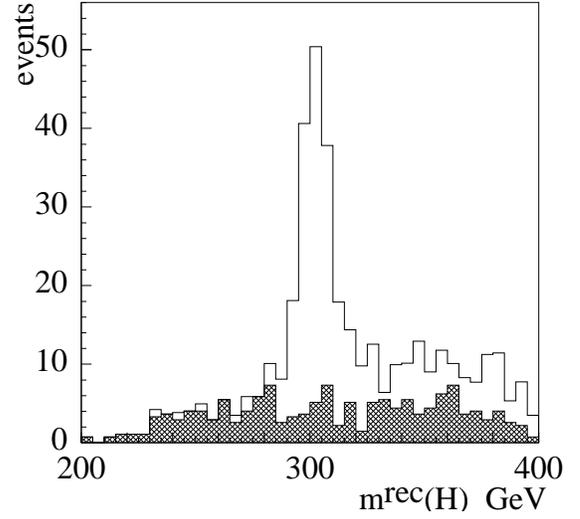}\\
\vspace*{-1cm}
\caption{Expected charged Higgs boson signal and background.}
\label{fig:hphm2}
\end{figure}

The discovery of charged Higgs bosons would immediately prove that
physics beyond the SM exists. 
The reaction 
$$\rm e^+e^-\rightarrow Z \rightarrow H^+H^-\rightarrow t\bar b\bar t b$$
can be observed at a LC~\cite{as:hphm500} and recent high-luminosity 
simulations~\cite{hphm} show that the production cross section times
branching ratio can be measured very precisely:
$$\Delta(\sigma BR({\rm H^+\rightarrow t\bar b}))/
        \sigma BR({\rm H^+\rightarrow t\bar b})=8.8\%$$
for $m_{\rm H^\pm}=300$~GeV, $\sqrt s = 800 $~GeV and 
${\cal L}=1000$~fb$^{-1}$.
A detailed reconstruction of the entire decay chain, as
illustrated in Fig.~\ref{fig:hphm}, is possible.
Figure~\ref{fig:hphm2} shows a clear expected charged Higgs boson
signal and small background.

\subsection{Determination of \boldmath$\tan\beta$\unboldmath}
\begin{figure}[htb]
\includegraphics[width=\columnwidth]{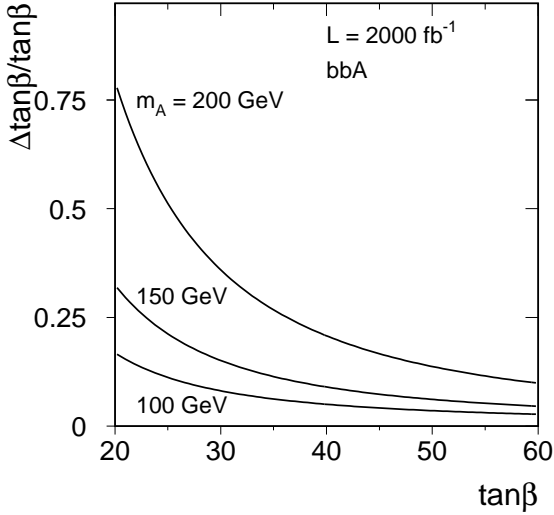}
\vspace*{-2cm}
\caption{2HDM $\tan\beta$ determination.}
\label{fig:tgb}
\end{figure}

The ratio of the vacuum expectation values
$\tan\beta$ can be measured with several methods.
The pseudoscalar Higgs boson, A, could be produced
via radiation off a pair of b-quarks:
$$\rm e^+e^-\rightarrow b\bar b\rightarrow b\bar b A \rightarrow b\bar b b\bar b.$$
Figure~\ref{fig:bba} illustrates the production process. 
The $\rm b\bar bA$ coupling is proportional to $\tan\beta$ and thus 
the expected
production rate is proportional to $\tan^2\beta$.
A precision better than 10\% can be achieved for a
Higgs boson mass of 100~GeV and large $\tan\beta$ values~\cite{gunion}. 
The sensitivity decreases with increasing Higgs boson masses and decreasing 
$\tan\beta$ values as shown in Fig.~\ref{fig:tgb}. 
This study assumes a luminosity of 2000~fb$^{-1}$, corresponding 
to several years of data taking.

\begin{figure}[htb]
\begin{Feynman}{80,40}{0,27}{0.8}
\put(25,40){\fermionul}        \put(5,22){${\rm e^-}$}
\put(25,40){\fermionur}        \put(5,55){${\rm e^+}$}
\put(15,30){\photonrightthalf} \put(32,30){$\gamma$}
\put(25,40){\gaugebosonright}  \put(35,43){${\rm \gamma/Z}$}
\put(55,40){\gaugebosonurhalf} \put(72,55){${\rm b}$}
\put(55,40){\gaugebosondrhalf} \put(72,22){${\rm \bar b}$}
\put(65,50){\gaugebosondrhalf} \put(80,32){$\rm A $}
\end{Feynman}
\vspace*{-0.5cm}
\caption{bbA production to determine the value of $\tan\beta$.}
\label{fig:bba}
\end{figure}

There are further methods to determine $\tan\beta$:
\begin{itemize}
\item The $\rm b\bar b\bar b b$ rate 
      from the pair-production of the heavier scalar in association 
      with the pseudoscalar Higgs boson,
      $$\rm e^+e^- \rightarrow HA \rightarrow b\bar b b\bar b,$$
      can be exploited.
      While the HA production rate is almost independent of $\tan\beta$ 
      the sensitivity is achieved owing to the variation of the 
      decay branching ratios with $\tan\beta$. 
\item The value of $\tan\beta$ can also be determined from the 
      H and A decay widths, which can be obtained from the previously
      described reaction.
\item The pair-production
rate and total decay width of charged Higgs bosons
      can contribute to the determination of $\tan\beta$.
      Charged Higgs boson production can also be used at the LHC 
      to measure $\tan\beta$~\cite{assamagan}.
\end{itemize}

\begin{figure}[htb]
\includegraphics[width=0.8\columnwidth,angle=90]{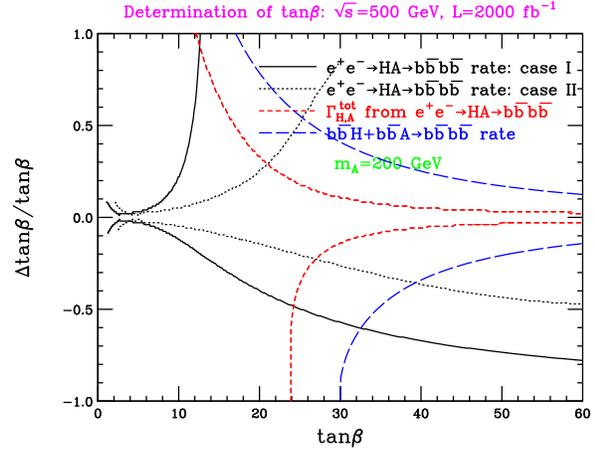}
\vspace*{-1cm}
\caption{MSSM $\tan\beta$ determination.}
\label{fig:mssmtgb}
\end{figure}

The results from the methods involving the neutral
Higgs bosons are summarized in Fig.~\ref{fig:mssmtgb}
for\,the MSSM.\,As the H and A decay rates depend on the MSSM parameters, two cases are
considered. In case~(I) heavy Supersymmetric particles are expected,
while in case~(II) the Higgs bosons could decay into light Supersymmetric
particles.

\subsection{Invisible Higgs boson decays}
\begin{figure}[htb]
\includegraphics[width=\columnwidth]{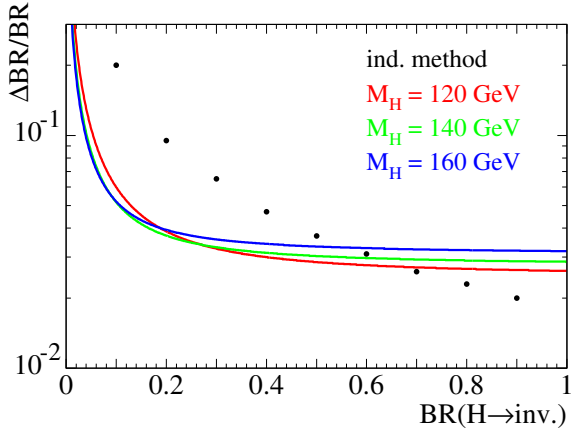}
\vspace*{-1.3cm}
\caption{Expected precision on the invisible Higgs boson 
         branching fraction.
         The dots represent the indirect method.}
\vspace*{-0.5cm}
\label{fig:inv}
\end{figure}

Among other possibilities, invisible Higgs boson decays could 
occur from the reaction involving neutralinos:
$$\rm e^+e^-\rightarrow ZH\rightarrow Z\tilde\chi^0\tilde\chi^0.$$ 
In this case the Higgs boson mass can be reconstructed from the recoiling
mass of the visible decay products:
$$m_{\rm H}=m_{\rm Z}^{\rm recoil}.$$
At LEP all Z decay modes involving charged fermions contributed to the
search,
while in a recent LC study~\cite{schumacher} so far only the hadronic 
decay mode $\rm Z\rightarrow q\bar q$ has been investigated.
This study for $\sqrt s = 350 $~GeV and ${\cal L}=500$~fb$^{-1}$
gives higher sensitivity compared to indirect methods
($1-$ sum of visible H decay modes).
For Higgs boson branching ratios $BR_{\rm i}$ into invisible decay 
products larger than 20\% and a SM Higgs boson production rate,
$$\Delta BR_{\rm i} /BR_{\rm i} <4\%$$ can be achieved.
Figure~\ref{fig:inv} shows the resulting sensitivities as a function
of the branching ratio into invisible decays.
\clearpage

\subsection{Higgs boson parity}

After the discovery of one or several Higgs bosons,
it is very important to determine the parity of the Higgs bosons and
distinguish a CP-even H boson from a CP-odd A boson.
This could be achieved by investigating the Higgs boson decay 
properties into $\tau$-leptons. 
The subsequent decay of the $\tau$'s into $\rho$'s and pions 
through the reaction
$$\rm H/A\rightarrow \tau^+\tau^-
\rightarrow\rho^+\bar\nu_\tau\rho^-\nu_\tau
\rightarrow \pi^+\pi^0\bar\nu_\tau\pi^-\pi^0\nu_\tau$$
 is studied~\cite{bower}.

\begin{figure}[htb]
\vspace*{0.8cm}
\includegraphics[width=0.8\columnwidth]{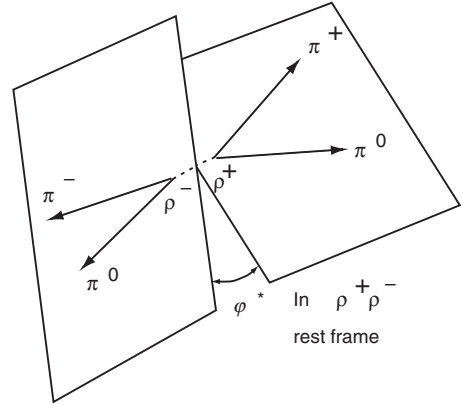}
\vspace*{-2cm}
\caption{Acoplanarity angle definition.}
\vspace*{-0.5cm}
\label{fig:def}
\end{figure}

\begin{figure}[htb]
\includegraphics[width=\columnwidth]{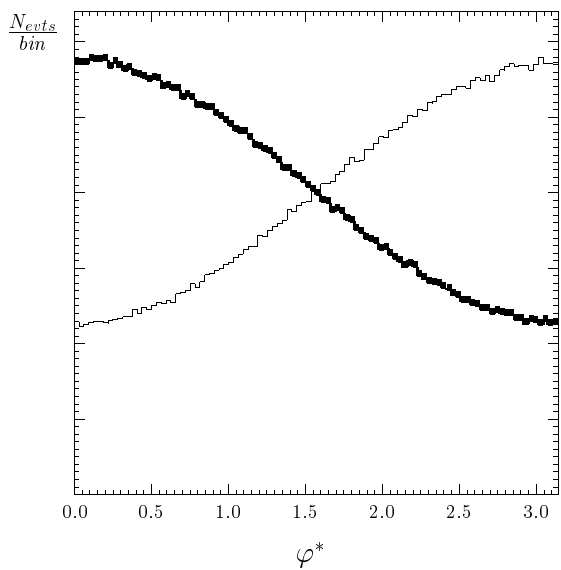}
\vspace*{-1cm}
\caption{Higgs boson parity determination before detector simulation.}
\vspace*{-0.5cm}
\label{fig:parity}
\end{figure}

\begin{figure}[htb]
\includegraphics[width=\columnwidth]{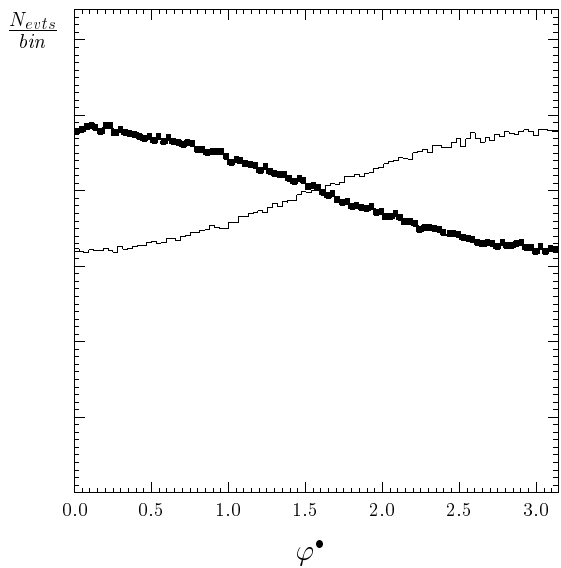}
\vspace*{-1cm}
\caption{Higgs boson parity determination after a preliminary
         detector simulation.}
\label{fig:parity2}
\vspace*{-0.5cm}
\end{figure}

The $\rho^+\rho^-$ acoplanarity angle is defined by the planes
of the pions in the rest frame of the $\rho$'s (Fig.~\ref{fig:def}). 
The acoplanarity angle distribution clearly distinguishes the different parity
states. Figure~\ref{fig:parity} shows the acoplanarity angle 
before detector effects are included, and Fig.~\ref{fig:parity2} after
a preliminary detector simulation.
The thick line is the expectation for scalar Higgs bosons and the 
thin line for pseudoscalar Higgs bosons.

\subsection{Distinction of Higgs boson models}
\begin{figure}[htb]
\includegraphics[width=\columnwidth]{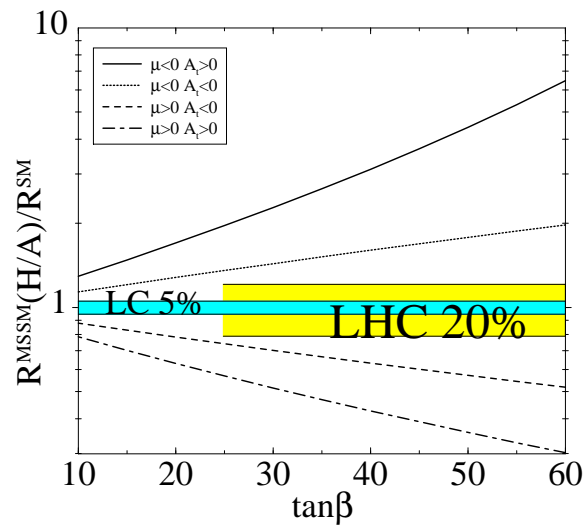} \\
\vspace*{-1cm}
\caption{Distinction of Higgs boson models from precision
         Higgs boson decay branching ratio determinations.}
\label{fig:models}
\end{figure}

The distinction of Higgs boson models is very important and could be based
on precision branching ratio measurements. 
The ratio of the Higgs boson decay rates into b-quarks and $\tau$-leptons
is defined by
$$R\equiv BR({\rm H/A\rightarrow b\bar b)}/
          BR({\rm H/A\rightarrow \tau^+\tau^-}).$$
The normalized value of $R$ to the SM expectation can be used 
to distinguish a general 2HDM from the MSSM. 
Large deviations from $R=1$ are
expected in the MSSM for several MSSM parameter combinations~\cite{guasch}
as shown in Fig.~\ref{fig:models}.
In relation to the LHC, where models can only be distinguished for 
$\tan\beta>25$, a LC covers the entire $\tan\beta$ range.

\begin{figure}[htb]
\includegraphics[width=\columnwidth]{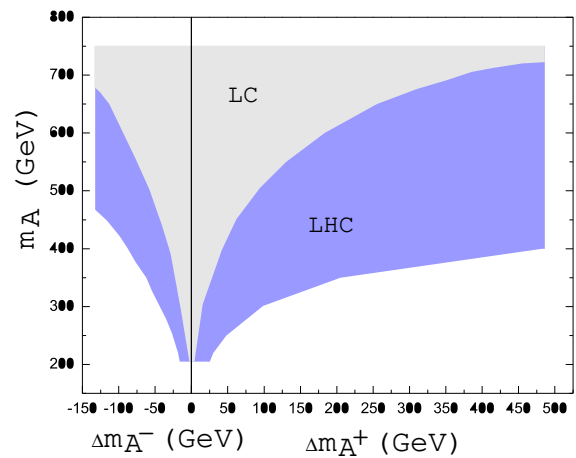} \\
\vspace*{-1cm}
\caption{Comparison of LHC and LC A mass predictions from precision
         scalar Higgs boson decay branching ratio determinations.}
\label{fig:lhclc}
\end{figure}

Important predictions on the pseudoscalar Higgs boson mass
can be made from the precision measurements of the scalar Higgs boson
branching ratios:
$$BR({\rm H\rightarrow b\bar b}) / 
  BR({\rm H\rightarrow WW}).$$
A LC obtains a precision normalized to the SM
expectation of less than $3.5$\%, while 
the expected precision at the LHC is less than $20$\%.
For a 400~GeV pseudoscalar Higgs boson A, its mass can be predicted 
at a LC with an error of about 50~GeV, while at the LHC only 
a lower mass limit can be set~\cite{gross}.
Figure~\ref{fig:lhclc} shows the sensitivity on the prediction of
the A boson mass at the LHC and a LC for a wide range of A masses. 

Beyond the MSSM, the Higgs boson particle spectrum is enriched
for example in the framework of the Non-Minimal Supersymmetric extension
of the SM (NMSSM)~\cite{miller} where an
extra Higgs boson singlet $\lambda NH_1H_2$ is present.
Such a model could be distinguished from the MSSM by precision
measurements of the Higgs boson masses and comparison with the 
predictions. 
Moreover, additional light neutral Higgs bosons may be observed
and the mass sum rules are modified, leading, for example,
to a reduced mass of the charged Higgs boson.

\clearpage
\section{CONCLUSIONS}

\begin{itemize}
\item After a first discovery and initial
      precision measurements in some decay modes
      at the Tevatron or the LHC,
      already in the first phase of a LC,
      many Higgs boson decay modes will be measured 
      with very high precision.
\item The precise LC data will allow the determination of the nature of the
      Higgs sector. 
      Models like the SM, the general 2HDM, the MSSM and the NMSSM 
      will be distinguished for a wide range of parameters.
\item The underlying mechanism of symmetry breaking and 
      mass generation will be tested.
\item Like for the top quark (LEP mass prediction, Tevatron observation),
      important consistencies of the model can be probed with
      combined LC and LHC physics. 
\item After 10 years of preparatory studies the LC has a 
      solid case and
      the high-energy physics community is prepared to 
      answer fundamental questions
      over the coming decades.
\end{itemize}  

\vspace*{5mm}
\section*{ACKNOWLEDGEMENTS}
I would like to thank 
M.~Battaglia,
J.-C. Brient,
K.~Desch,
F.~Gianotti,
A.~Gay,
E.\,Gross, J.\,Guasch, A.\,Kiiskinen, R.\,Van Kooten,
N.\,Meyer, D.\,Miller, S.\,Pe\~naranda, J.\,Schreiber,
M.~Schumacher, S.~Yamashita, and P.~Zerwas 
for help with the preparation of this presentation.

\end{document}